\documentclass[preprint,12pt]{elsarticle}

\usepackage{amssymb}

\usepackage[utf8]{inputenc}
\usepackage{amsmath}        
\usepackage{amsfonts}       
\usepackage{amsthm}         
\usepackage{bbding}         
\usepackage{bm}             
\usepackage{graphicx}       
\usepackage{fancyvrb}       
\usepackage{natbib}         
\usepackage{dcolumn}        
\usepackage{booktabs}       
\usepackage{paralist}       
\usepackage[usenames]{xcolor}  
\usepackage{float}
\restylefloat{table}
\usepackage{tabulary}
\usepackage{longtable}
\usepackage{wrapfig}
\usepackage{caption}
\usepackage{hyphenat}
\usepackage{url}
\usepackage{multirow}
\usepackage{color}
\usepackage{rotating}
\usepackage{lscape}
\usepackage[ruled,vlined]{algorithm2e}

\theoremstyle{plain}

\numberwithin{equation}{section}

\journal{Applied Soft Computing}

\begin{document}

\begin{frontmatter}



\title{Using CMA-ES for Tuning Coupled PID Controllers within Models of Combustion Engines}


\author{Kate\v{r}ina Henclov\'{a} \footnote{Email address: henclova@karlin.mff.cuni.cz}}
\address{Regional Innovation Centre for Electrical Engineering,\break University of West Bohemia, \break Univerzitn\'{i} 26, 306 14 Pilsen, \break Czech Republic}

\begin{abstract}
Proportional integral derivative (PID) controllers are important and widely used tools in system control. Tuning of the controller gains is a laborious task, especially for complex systems such as combustion engines. To minimize the time of an engineer for tuning of the gains in a simulation software, we propose to formulate a part of the problem as a black-box optimization task. 

In this paper, we summarize the properties and practical limitations of tuning of the gains in this particular application. We investigate the latest methods of black-box optimization and conclude that the Covariance Matrix Adaptation Evolution Strategy (CMA-ES) with bi-population restart strategy, elitist parent selection and active covariance matrix adaptation is best suited for this task. 
Details of the algorithm's experiment-based calibration are explained as well as derivation of a suitable objective function. The method's performance is compared with that of PSO and SHADE. Finally, its usability is verified on six models of real engines.
\end{abstract}


\end{frontmatter}




\section{Introduction}

In a running combustion engine, one or more PID controllers ensure that certain quantities (e.g. intake pressure or exhaust gas temperature) remain constant or within given range. Since these quantities may be naturally related and affect each other, controllers are often coupled and their gains cannot be tuned independently of each other.

When engines are modeled, as in simulation software WAVE\footnote{WAVE is 1D engine and gas dynamics simulation software package developed by Ricardo Software [\citenum{WAVEmanual}] and used by the author of this paper.}, controllers may and need to be tuned using simulations with little or no knowledge about the transfer function that describes the system. The complicated finite-element model cannot be described by a simple formula and, moreover, it comes with its own modeling and discretization errors.

Presently, manual work makes up a major part of the controller tuning process. This lengthy procedure is based on trial and error and requires a knowledgeable and experienced control engineer. For systems with a single controller (or multiple but decoupled controllers), simple rules of thumb can be employed (e.g. Ziegler-Nichols [\citenum{ZNmethod}]). Similar, already-solved problems can also provide a guideline. However, when having a complicated or unique system of coupled controllers, the complexity of the task makes it very difficult to solve even for an experienced control engineer. Moreover, in our application of PID controllers within combustion engine models, other professionals need to tune the controllers as well, creating the need for a simple-to-use, robust tool.

Our goal is to formulate at least a part of the problem and deliver a method that would eliminate or significantly lower the need for manual tuning. It should find a solution of a given problem within acceptable time and with as little user interaction as possible. When combined with simple tuning rules or educated guesses, our method is to use the provided solution approximation as a starting point and quickly find a more refined solution. 
Further testing of solutions found by our method (such as verifying their robustness) is, however, still left for manual post-processing. In principle, it is simple to replace the proposed objective function by its expected value over parametric uncertainties. However, the uncertainties on the parametric space are not available for the considered application and thus it is not a part of this work. Moreover, such approach would significantly increase the computational load.

The proposed method is used to design the controller gains for a predefined trajectory of the controller. For more complex system that requires the use of gain scheduling [\citenum{aastrom2013adaptive}] or fuzzy PID [\citenum{petrov2002fuzzy}], the proposed method will be used to design the gain in the operational point designed by the supervising engineer. The engineer will also be responsible for selection of the gain scheduling or fuzzy rules.

The PID tuning problem with either one controller or multiple but decoupled or symmetric controllers can be and has been reformulated as a black-box optimization problem and solved with an appropriate method. Evolutionary algorithms have been successfully applied to many engineering problems [\citenum{metaheuristics}] and they have also been used to tune PID controllers. The genetic algorithm [\citenum{goldberg1, holland1}] was one of the first to be used [\citenum{PIDwithGAold}], and many others followed such as the differential evolution (DE) [\citenum{de2,de3,de1}] in [\citenum{PIDwithDE, Baskar2009}] or the particle swarm optimization (PSO) [\citenum{swarm1}] and its combinations (e.g. with bacterial foraging algorithm) in [\citenum{PIDwithPSO, PIDwithPSO_2, Baskar2009, BF-PSO_tuning}].

The tuning problem with multiple coupled controllers can be formulated as an optimization problem as well but it adds another level of complexity to the task. (To the author's knowledge, there have not been any papers dealing with such problems.) Even though there are naturally multiple objectives (one for each controlled quantity), we can use a simple trick to carefully transform them into a single objective and thus enable the problem to be solvable by usual means. This way, however, the problem becomes harder to solve, while the time budget remains small. With simulations taking up to several minutes each, we aim for an overnight or a one-day computation on a regular PC, i.e. a few thousand simulation runs at most. This imposes high expectations upon efficiency of the method used.

Considering properties of the problem, we choose to use a variant of the Covariance Matrix Adaptation Evolution Strategy (CMA-ES) [\citenum{hansenDerandomized, EStheory2015, bipop, activecma}], an evolutionary algorithm founded deep in probability theory. It has proven to be very effective and robust method in the extensive testing of Black-Box Optimization Benchmarking (e.g. [\citenum{BBOB2009, BBOB2010}]), surpassing the above mentioned algorithms and many others (on the relevant sort of problems). Despite its fame in the optimization community and large number of practical applications, it has so far been little used for tuning PID controllers [\citenum{Baskar2009, Baskar2010, PIDwithCMAES}] or similar problems [\citenum{hansen2009combustion,hansen2008thermoacoustic}].

In this paper, we experiment with the algorithm's settings and come up with a method that meets practical requirements, such as tolerable runtime, and is thus fit for common use by engineers working with complicated simulations (e.g. users of WAVE). Moreover, we demonstrate its applicability on six models of real-world engines with one, two and three controllers.

For comparison with CMA-ES, we use PSO (as implemented in the software package DEAP [\citenum{deap}]) and a restarted version of SHADE (Success-History based Adaptative DE [\citenum{shade}]) with an implementation based on that by Tanabe [\citenum{shade_implementation}]. We show that they do not perform by far so well as CMA-ES.


\section{Formulation of the problem} \label{s:problem}

In this section, we formulate the controller tuning problem as a problem of numerical optimization. This part is essential as it determines whether the problem will be solvable (with the given practical limits) and what method should be used. Compared to other research on controller tuning [\citenum{PIDwithDE, PIDwithPSO, PIDwithPSO_2, Baskar2009, BF-PSO_tuning, PIDwithGAold}], dealing with coupled controllers requires an extra level of complexity.

\subsection{PID controllers in the context of engine simulations} \label{s:PIDs}

PID controllers are well known and powerful tools in system control [\citenum{controlSystemDesign,Ogata}]. To describe their inner workings, we will use the traditional, though perhaps old-fashioned, notation. A controller's input is the error function 
\begin{equation} \label{eq:errorfun}
e(t) = \left| actual(t) - target(t) \right|,
\end{equation}
i.e. the time-dependent absolute difference between the desired target value and the actual value of a quantity (as measured by a sensor or computed by a model). The output control signal that defines the system's subsequent reaction is given as
\begin{equation} \label{eq:control}
C(t) = P e(t) + I \int_{0}^{t} e(\tau) d\tau + D\ \frac{d}{dt} e(t),
\end{equation}
where $P$, $I$ and $D$ are the proportional, integral and derivative gains, respectively.

The controllers' implementation is provided within the simulation software WAVE. All the computations as well as the whole engine model are given to us as a black box and we do not analyze them. The software as well as the models are carefully calibrated to comply with the real world. The simulation is fully deterministic and we do not introduce any random perturbations of environment variables during the tuning process.

We input the controllers' gains into the simulation software and obtain computed trajectory of the controlled quantities over time. Our goal is to find such \emph{constant} gains $P$, $I$ and $D$ for each controller, that the corresponding controlled quantities converge to the target value and do so as quickly as possible. This provides us with the key how to compare solutions' quality. 

The entire tuning process is offline but, due to nonlinearities in the system, the curve is not as smooth as it is common for simple systems.

In this work, we are always provided with initial values and with (nonzero) constant target values of the controlled quantities. The values are given by the context of combustion engines.


\subsection{The optimization problem}

With $k$ controllers within a system, each determined by three constant gains $P$, $I$ and $D$, there are $3k$ gains to be tuned: 
$
x = ( P_1, I_1, D_1,\ \ldots \ ,\  P_k, I_k,$ $ D_k ).
$
Let us further suppose that we are given an objective function $M: \mathbb{R}^{3k} \rightarrow \mathbb{R}$. It returns a single number for every vector of controllers' gains and describes how well the whole system is controlled by controllers with such gains. Assuming that higher quality inputs have lower function values, we search for the objective function's minimizer. The numerical optimization problem derived from the controller tuning problem can be formulated as looking for $x^*$ such that
$
M(x^*) = \mathrm{ess inf}_{x \in \mathbb{R}^{3k}} M(x).
$
I.e. $x^*$ is a minimizer of $M$ that is reachable by numerical means. Naturally, we do not strive to find the perfect optimizer $x^*$ but rather any good-enough solution $x^{**}$ for which it holds 
$
M(x^{**}) - M(x^{*}) < \epsilon,
$
where $\epsilon > 0$ is a given tolerance. Such ``minimizer'' $x^{**}$ is not unique but, for practical purposes, it need not be.

In our case, function $M$ must incorporate a finite-element model of an engine and process its results. The model's input is the vector of controllers' gains (a candidate solution) and it outputs development of the controlled quantities' values over time (the controllers are a part of the model). These outputs then need to be processed in order to obtain a single number. Here, it is essential to carefully encode information about the candidate solutions' quality, so that their final values rank them well.



\subsection{Definition of the objective function} \label{s:objfun}

When the controllers' gains are set and the simulation is run, it outputs the error functions 
$
e_i(t) = e_i(x,t),\ i = 1, \ldots, n,
$ 
described by \eqref{eq:errorfun}. We shall now focus on how to process $e_i$ so that the final function value contains all information about quality of the input. 
In this section, the vector of controllers' gains $x$ is arbitrary but fixed. So, for simplicity of notation, it is left out from the functions' arguments: we write $e_i(t)$ instead of $e_i(x,t)$ etc.


\subsubsection{Controlling multiple objectives}

Let us assume for a moment that once controllers' gains are set, quality of the setting is characterized by functions 
$
F_i(t): \mathbb{R}^{3k} \rightarrow \mathbb{R},\ i=1,\ldots,n,
$
that describe the $n$ controlled quantities' development (simulation time $t$ being given). That provides us with $n$ objectives to be minimized. In our case, the number of controllers is the same as the number of controlled quantities (i.e. $k = n$) but, in general, this is not required. We expect $k$ to be small, typically $k \leq 5$.

We shall combine these objectives into a single one. Using a weighted sum (see e.g. [\citenum{marler2004survey}]), we can define the objective function describing the response of the whole model as
$
F(t)\ =\ \sum_{i = 1}^{n} \ w_i\ F_i (t),
$
where $w_i$ is the weight constant corresponding to the $i$-th controlled quantity. The purpose of constants $w_i$ is to express our priorities and, more importantly, make all the objective functions $F_i$ comparable, as they may significantly differ in range depending on the corresponding units. Therefore, we set
$
w_i = p_i / scale_i,
$
where $p_i$ describes priority of the $i$-th objective and $scale_i$ contains knowledge of its typical value or range. In our case, we set 
$
scale_i = target_i > 0
$
to be the (constant and nonzero) target value of the $i$-th controlled quantity.
The $target_i$ element is the remainder of the integral under its (constant) curve, considering that all the objectives are measured for the same length of time.
In our experiments, we shall always set priorities 
$
p_i = 1,\ i=1,\ldots,n.
$

Thus we have obtained the way of combining multiple objectives into one:
\begin{equation} \label{eq:objfun_sum}
F(t)\ =\ \sum_{i = 1}^{n} \ \frac{1}{|target_i|} F_i (t).
\end{equation}


\subsubsection{Step-response based objective function}

\begin{figure}
\centering
\includegraphics[width=8cm]{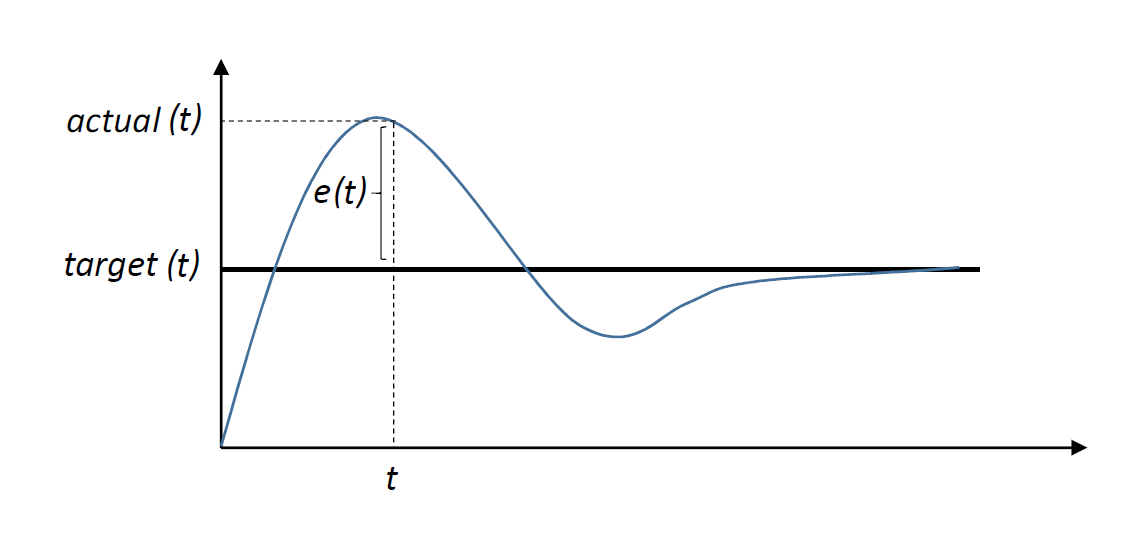}
\caption[]{For a given setting of controllers, the error at time $t$ is given as $e(t)=\left| actual(t) - target(t) \right|$.}
\label{fig:errorT}
\end{figure}

It remains to define $F_i$, the function corresponding to the $i$-th objective (controlled quantity), $i=1,\ldots,n$. The general framework is based on how the step response looks like, see the idealized step response in figure \ref{fig:errorT}. 

The step response describes how the controlled quantity behaves after an engine is started. The abrupt step change of the target value is the most difficult part for the controllers to deal with. Once the engine is running, the changes in the environment and the target values are not so sudden. Thus, when we tune the controllers for the ``worst case scenario'', there is a good reason to believe (supported by empirical evidence) that they will be able to control other cases as well. When the environment changes too much, e.g. hitting the extremes of RPM (rotations per minute), the controller gains may be changed using gain scheduling [\citenum{aastrom2013adaptive}].

Functions $F_i$ are naturally given by formula
\begin{equation} \label{eq:objfun_i}
F_i (t)\ =\ \int_{t_0}^{t}\ T(\tau)\ E(e_i(\tau))\ d\tau, \ \ \ 
t_0 \geq 0,
\end{equation}
where $t_0$ is the initial time (typically $t_0 = 0$), $T(\tau)$ is a function of time, $E(e_i(\tau))$ is a function of error, $e_i(\tau)$ is the error as defined in \eqref{eq:errorfun}. In our case, $t$ is given by the end of the simulation.

Function $E(e_i(\tau))$ describes our concern over the actual error \emph{size}. On the other hand, function $T(\tau)$ characterizes how much we care about \emph{when} the particular error occurs. Its value at a given time can also be perceived as weight assigned to the corresponding error function value. 
There are multiple commonly used criteria of step response quality (IAE, ITAE, ITSE or ISE), which all fit the framework described above, giving us common choices of functions $E(e_i(\tau))$ and $T(\tau)$. The well known ITAE criterion \cite{superITAE}
\begin{equation}\label{eq:itae}
ITAE (t)\ =\ \int_{0}^{t}\ \tau | e(\tau) |\ d\tau,
\end{equation}
fits our purpose the best, as we are highly concerned about when the error happens: it is not so important in the beginning, while at the end, error means nonconvergence. 

This information must not get lost even after combining multiple objectives in \eqref{eq:objfun_sum}. Therefore, to sort the candidate solutions better, we slightly modify \eqref{eq:itae}. First, we shift the time, so that the value of one second loses its importance as a factor. Then, we can choose to penalize nonconvergent solutions. We can choose to stress the importance of time by increasing the degree of its polynomial from linear to quadratic. Or, instead, we can leave out the beginning of the time interval as the large error that can be expected there is -- from the optimization viewpoint -- only noise. Various possibilities have been experimented with and the best option found was to set
\begin{equation} \label{eq:objfun_SITAE}
F_i(t)\ =\ \int_{t_0}^{t}\ (\tau+1) | e_i(\tau) |\ d\tau \ \ \ 
t_0 \geq 0.
\end{equation}
Here, the beginning of integration may be shifted from $0$ to arbitrary $t_0$. Selection of the shift $t_0$ along with its justification and influence upon performance is discussed later. The simulation length $t$ is set appropriately and is same for all controlled quantities.


\subsubsection{The objective function}

In previous steps, we have constructed the objective function

\begin{equation} \label{eq:objfun_FINAL}
M(x) = F(x,t)\ =\ \sum_{i = 1}^{n} \ \frac{1}{|target_i|} \int_{t_0}^{t}\ (\tau+1) | e_i(x,\tau) |\ d\tau,
\end{equation}
where $n, \ target_i$ and $0 \leq t_0 < t$ are given and the error functions $e_i$ are computed by the simulation software.


\subsection{Character of the problem and practical limitations}

Character of the problem plays a crucial role when choosing a fitting method to solve it. 
For any set of controllers' gains, the simulation provides us with all data necessary to compute the objective function value (equation \eqref{eq:objfun_FINAL}). However, we cannot analyze the model itself and that forces us to take it as a black box.
We can only make a few careful assumptions. In general, objective function $M$ is non-convex, non-linear, non-quadratic and highly multimodal (i.e. having multiple local optima). It is probably continuous, but we know nothing of its conditioning (i.e. when a function is ill-conditioned, it may be very steep in some places), and there are no derivatives available. We cannot presume that $M$ is smooth, so neither the derivative approximations would necessarily be sensible. However, we do not expect $M$ to be noisy due to its heart made of a FEM model. We also know that the objective function is non-separable, i.e. we cannot tune one parameter independently of the others and expect to find the optimum that way. One controller's parameter's value strongly affects other parameters and one controller's setting may affect the behavior of other controllers (the controllers may be coupled). Since the relationships between the variables are far from random, we would like the algorithm to mine and use this information to improve its search.

Beside this, we have to consider other practical limitations. Engine simulations, as in WAVE, take up to several minutes (per case -- when parameters like rotations per minute etc. are fixed), but the time budget is limited. Usually aiming for an overnight computation, we are allowed only a few thousand evaluations of the objective function. That is, when parallelization is used, and therefore we require the optimization algorithm to be parallelizable. Last but not least, the algorithm must be robust -- it must reliably produce decent results, even at the cost of somewhat slower performance.






\section{The optimization method} \label{s:cmaes}

Metaheuristic and evolutionary methods have been extremely successful when tackling hard black-box optimization problems [\citenum{metaheuristics}]. 
In order to choose a method appropriate for our problem, we consider the extensive Black-Box Optimization Benchmarking (BBOB) results, e.g. [\citenum{BBOB2009,BBOB2009tables,BBOB2010, COCO}], which compare many algorithms on a wide range of test functions. Following the relevant data -- tests on multimodal, non-separable, noiseless functions -- we choose the Covariance Matrix Adaptation Evolution Strategy with bi-population restart scheme (BIPOP-CMA-ES). Despite its efficiency and growing number of practical applications, its use for tuning of PID controllers is scarce [\citenum{PIDwithCMAES,Baskar2009, Baskar2010}].


\subsection{CMA-ES: idea of the basic algorithm}

The Covariance Matrix Adaptation Evolution Strategy (CMA-ES) [\citenum{hansenDerandomized,hansen2004multimodal}] is an evolutionary algorithm that uses stochastic and algebraic tools to define optimally diverse population of candidate solutions in an area that seems to be most promising. The size of the area and its location are determined based on the algorithm’s previous experience with the objective function. 
New candidate solutions are sampled from a multivariate normal distribution:
\begin{equation}
x_k \sim m + \sigma \mathcal{N}(0,C),\ \ k=1,\ldots,\lambda,
\end{equation}
whose mean $m$ and covariance matrix $C$ are adapted in each generation along with the general step size $\sigma$. The number of sampled candidate solutions $\lambda$ is called the population size.

The mean in the new generation is defined as weighted average of several best-ranking individuals (the parent set) of the last generation. The weights are constant and depend only on the individuals' relative ranking. The covariance matrix is supposed to capture information about the objective function's features (curvature) around the mean. Detailed description of the sophisticated updating of the covariance matrix can be found in the original paper [\citenum{hansenDerandomized}] or the informal tutorial [\citenum{CMAEStutorial}]. In short, the matrix of the previous generation is weighted and further improved by exploiting the maximum likelihood principle (when estimating the distribution's parameters from the data) and by adding information of the overall progress across generations (the evolution path). The latter technique is also used for controlling the overall step length $\sigma$.

Authors of CMA-ES have designed the method so that minimal user interaction or manual setting of parameters is necessary. The starting point is needed in order to initiate the first mean $m^{(0)}$ but otherwise all the parameters are assigned usable default values. A user might want to change only the initial step length $\sigma^{(0)}$ and the population size $\lambda$. The parent set size is given as 
$
\mu = \frac{1}{2} \lambda.
$
The population size is by default very small, given by the formula
$ 
\lambda_{\text{def}}\ =\ 4 + \left\lfloor 3 \log(d) \right\rfloor,
$
where $d$ is the problem dimension [\citenum{hansenDerandomized}]. For a given budget, smaller population size means faster adaptation. More often than not, this is a desirable property.
Choice of $m^{(0)}$ and $\sigma^{(0)}$ is described further.


\subsection{Extensions of the basic algorithm}

The basic algorithm described above can be further extended and upgraded. In our application, we shall use the elitist BIPOP-aCMA-ES version, i.e. Covariance Matrix Adaptation Evolution Strategy [\citenum{hansenDerandomized}] with active covariance matrix updates [\citenum{activecma}], elitist scheme of parent selection [\citenum{EStheory2015}] and bi-population restart strategy [\citenum{bipop}]. The choice of this particular variant, as well as setting the algorithm's parameters, is supported by numerical experiments described in section \ref{s:experiments}.


\subsubsection{Restart strategies}

Restart strategies enable the basic algorithm to become more robust by outweighing premature convergence. They are based on changing the population size, which also helps to balance exploration (exploring as much of the vector space as possible) and exploitation (thorough investigation of promising areas) better.

The default population size $\lambda_{\text{def}}$ is small and the optimal population size may be much larger. The IPOP (increasing population) strategy doubles the population size (the factor of $2$ is empirical) every time a new restart is launched, while other parameters remain unchanged [\citenum{ipop}].

The more advanced BIPOP (bi-population) restart strategy [\citenum{bipop}] makes use of two interlacing regimes. The first one uses the IPOP restart strategy, while the second uses varying small populations. After each restart, it is decided which of the two regimes is to be applied next depending on whose count of conducted function evaluations is lower. The maximal population size is limited by the number of restarts under the first regime.


There are two prominent restart criteria that need to be set carefully and adjusted for each model, because they depend on its numerical values. The first one is \texttt{TolFunHist}, which gives tolerance in function value history (range of the best objective function values of several last generation is almost zero). The second criterion is \texttt{TolFun}, which gives tolerance in function values of the current generation (stop, if the difference between the current best and worst candidate solution is \texttt{TolFun}-negligible) and it also checks the history of function values in a very similar way to \texttt{TolFunHist}. Their default values are too small for our use and better values can be guessed from the objective function's numerical values.

It was observed that it is generally good to try larger values of \texttt{TolFunHist} first and diminish it if it invokes restarts too often. When logging the algorithm's run, occurrence of premature restarts is usually obvious. 
Criterion \texttt{TolFun} can save some runtime too, when it is set up correctly (in the same way as the first criterion). Or we can set it to a small value, preferably a fraction of \texttt{TolFunHist}. That way, most restarts are called by \texttt{TolFunHist} and we need to tune only this one restart criterion.


\subsubsection{Elitist selection} \label{ss:elitist}

There are multiple options of how to choose parents of a new generation [\citenum{EStheory2015}]. The basic algorithm selects all parents from the current generation (non-elitist selection). However, this scheme does not exploit very good solutions found early in the search. They are used as parents within the one generation that produced them and then are lost. Selecting the parent set among the individuals of the current generation and their parents as well (elitist selection) helps to preserve the exceptionally good individuals until they are superseded and thus amplify their influence. This approach speeds up the convergence and is advantageous in our problem, where good solutions are scarce. The downside is that it may, and sometimes it does, lead to premature convergence to a local optimum.


\subsubsection{Active covariance matrix adaptation} \label{ss:active}

In the original algorithm, the successful individuals are used for the covariance matrix adaptation. Variance in directions that have proven to be beneficial is increased and thus they are preferred when sampling next generation.
With the so-called active updates, information hidden in the unsuccessful individuals is exploited as well [\citenum{activecma}]. As opposed to passive decay over time, variance in detrimental directions is actively decreased. Simply put, besides telling the method where to go, we also tell it where \emph{not} to go.


\subsection{Important properties of CMA-ES and their use}

CMA-ES does not use gradients and it does not even presume their existence. Moreover, it does not even use the actual values of the objective function once relative ranking has been assigned to the candidate solutions (except for some stopping criteria). As a result, it is invariant to strictly monotonic transformations of the objective function. I.e. those transformations of the objective function that have no effect upon the relative ranking of individuals do not effect the method's performance, making it more robust.

Further, the method exhibits invariance to invertible linear transformations of the search space. In particular, CMA-ES is invariant to scaling of variables (coordinate axes), which is the key property that makes it well-suited for tuning multiple controllers: parameters of one controller are usually of roughly the same scale, but with multiple controllers, the scaling may differ by many orders.
Scaling of variables is used in the following way. The algorithm is given an initial approximation of the solution (hereafter called the reference point): $s = ( s_1, \ \ldots, s_N )$. Then, when the algorithm wants to evaluate vector $v = ( v_1, \ \ldots, v_N )$, it scales it by the reference point's elements' magnitudes and inputs vector $w$ in the engine model: 
$
w = ( |s_1| \cdot v_1, \ |s_2| \cdot v_2, \ \ldots , \ |s_N| \cdot v_N )
$.
Because the scale coefficients are all positive and signs of the variables are essential, the algorithm must always have a zero vector as its starting point. Also, the initial step length is set to $\sigma^{(0)} = 1.0$.

{ 
  \scriptsize
	

\begin{algorithm}[H]
	\DontPrintSemicolon

  set $\lambda, \mu$ \;
	initialize $m, \sigma, C=I, p_{\sigma} = 0, p_c = 0$ \;
	initialize $restart\_regime = 1, count_1 = 0, count_2 = 0$ \;
  \;
	
	\While{termination criteria not met}{

		\While{restart criteria not met}{
		
		  \If{not first generation in a restart}{
				\For{$i = 1, \ldots, \mu $}{
					$x_{i+\mu} = x_i$                   \tcp*{relabel parents of previous generation}
					$f_{i+\mu} = f_i$ 									\tcp*{relabel parents' objective function values}
				}
		  }
			\;
	
			\For{$i = 1, \ldots, \lambda $}{			
				$x_i \sim \mathcal{N}(m,\sigma^2 C)$  \tcp*{sample new population from normal distribution} 
				$f_i = evaluate(x_i)$                 \tcp*{evaluate $x_i$ with objective function} 
			}			
			\;
			
			sort $x_i, i = 1, \ldots, \lambda+\mu$ acc. to $f_i$                             \tcp*{assign relative (descending) ranking}
			
			$m^* = m$ \;
			$m = update\_m (x_i, \ldots, x_{\mu})$                                             \tcp*{move the mean utilizing the parents}
			\tcp{the evolution paths contain information about past progress}
			$p_{\sigma} = update\_p_{\sigma} (p_{\sigma}, \sigma^{-1} C^{-1/2} (m-m^*))$       \tcp*{isotropic evolution path update}
			$p_c = update\_p_c (p_c, \sigma^{-1} (m-m^*), \|p_{\sigma} \| )$                   \tcp*{anisotropic evolution path update}
			$C = update\_C (C, p_c, (x_1 - m^*)/\sigma,\ldots,(x_{\lambda+\mu} - m^*)/\sigma)$ \tcp*{covariance matrix update}
			$\sigma = update\_\sigma (\sigma, \| p_{\sigma} \|) $                              \tcp*{step size update}
			
			\;
			
			\If{$restart\_regime = 1$}{
			  $count_1 = count_1 + \lambda$ \;
			}
			\Else
			{
			  $count_2 = count_2 + \lambda$ \;
			}
			
		}
		\;
		
		\If {$count_1 < count_2$} {
		   restart\_regime = 1 \;
		}
		\Else{
		   restart\_regime = 2 \;
		}
		\;
		
		reinitialize parameters and variables acc. to selected restart regime \;
		    
	}

\caption{Elitist BIPOP-aCMA-ES \label{alg:cmaes}}

\end{algorithm}

\vspace{20pt}

}


\section{Experiment-based calibration of the optimization method} \label{s:experiments}

For the experiments, the 1D engine simulation software package WAVE by Ricardo Software was used [\citenum{WAVEmanual}], as the primary incentive was to develop a working automated tuner of PID controllers within WAVE. For CMA-ES implementation, we take the Python code distributed by Hansen [\citenum{cmapy}].

Since models of engines take several minutes to run, while the CMA-ES runtime is negligible, we shall measure time in number of evaluations of the objective function ($=$ number of simulation runs).


\subsection{The basic testing model}

For the algorithms' calibration and basic testing, a quick-to-run model was used. It has three strongly coupled controllers and, despite its relatively simple mechanics, it proved hard to tune.

\begin{figure*}
\centering
\includegraphics[width=0.85\textwidth]{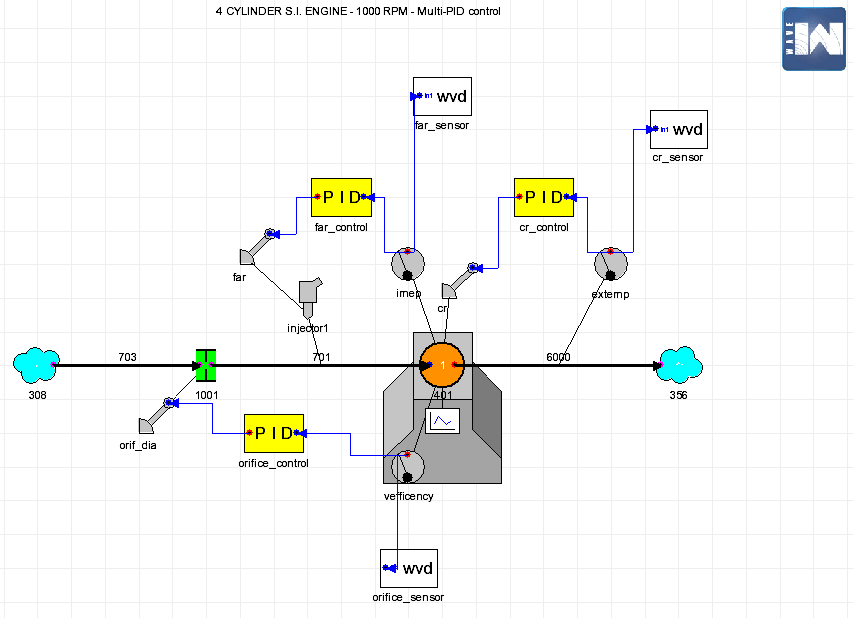}
\caption[The basic testing model in WAVE.]{The basic testing model in WAVE.}
\end{figure*}

In this model, we can see a single cylinder (orange circle) engine. The blue ``clouds'' contain information about the surroundings (e.g. ambient pressure and temperature or initial fluid composition). The thick black lines depict the ducts. The green element is an orifice -- an opening of variable diameter. The yellow PID elements are the controllers, whose gains we want to tune. From left to right, the first PID controls the orifice diameter, the second one controls the fuel-air ratio by manipulating the fuel injector, and the third controls the compression ratio (the ratio of the maximum to minimum volume in the cylinder). The arrow-like elements are actuators that perform the actual mechanical control based on the control signal outputted by the corresponding controllers.

The controlled quantities measured by sensors (depicted as gray circles) are: indicated mean effective pressure (IMEP; the average pressure acting upon the piston during its cycle; controlled by adjusting the fuel-air ratio), exhaust gas temperature (controlled by the compression ratio, i.e. the ratio of largest and smallest possible capacity of the combustion chamber), volumetric efficiency (the ratio of the volume of fluid actually displaced by a piston; controlled by opening of the orifice). Clearly, they have nontrivial influence on each other, so the controllers are coupled and cannot be tuned independently of each other.

\begin{figure}
\centering
\includegraphics[width=\textwidth]{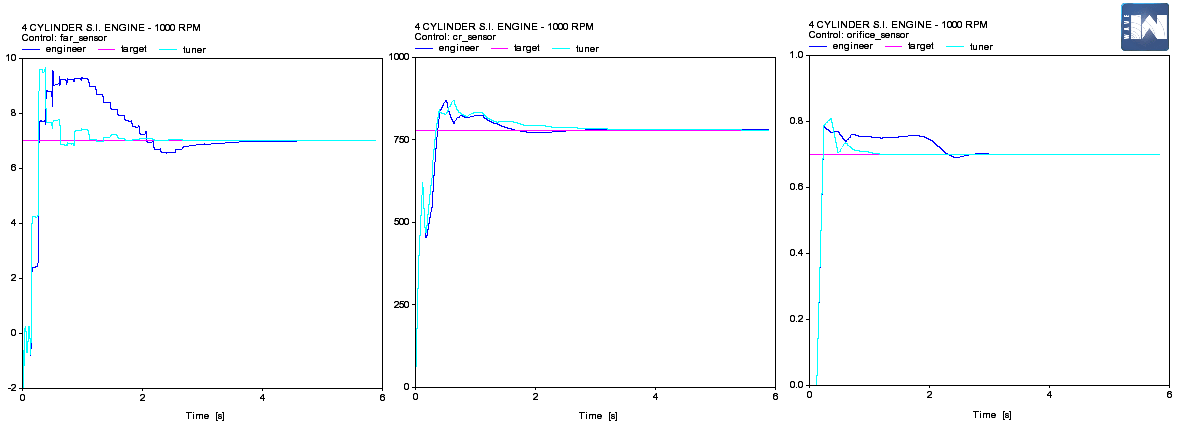}
\caption[The basic testing model: comparison of good solutions.]{The basic testing WAVE model: comparison of good solutions. Pink: the target value, dark blue: tuned by engineer, light blue: tuned by the algorithm.}
\label{fig:good}
\end{figure}

We then simulate the start of this engine, when the controllers face a simultaneous step change of all three target values. The target values of the controlled quantities are given to us (appropriately to the context). 
In figure \ref{fig:good}, each plot contains two step responses. Controllers tuned by an engineer (hereafter called the baseline solution) yield the dark blue curves that describe the development of the controlled quantities' values. This solution is sufficient but we can see that a solution found by our method (light blue) is even better. More importantly, it can be reached with very little user interaction, saving a lot of expert manual work.

\begin{figure}
\centering
\includegraphics[width=\textwidth]{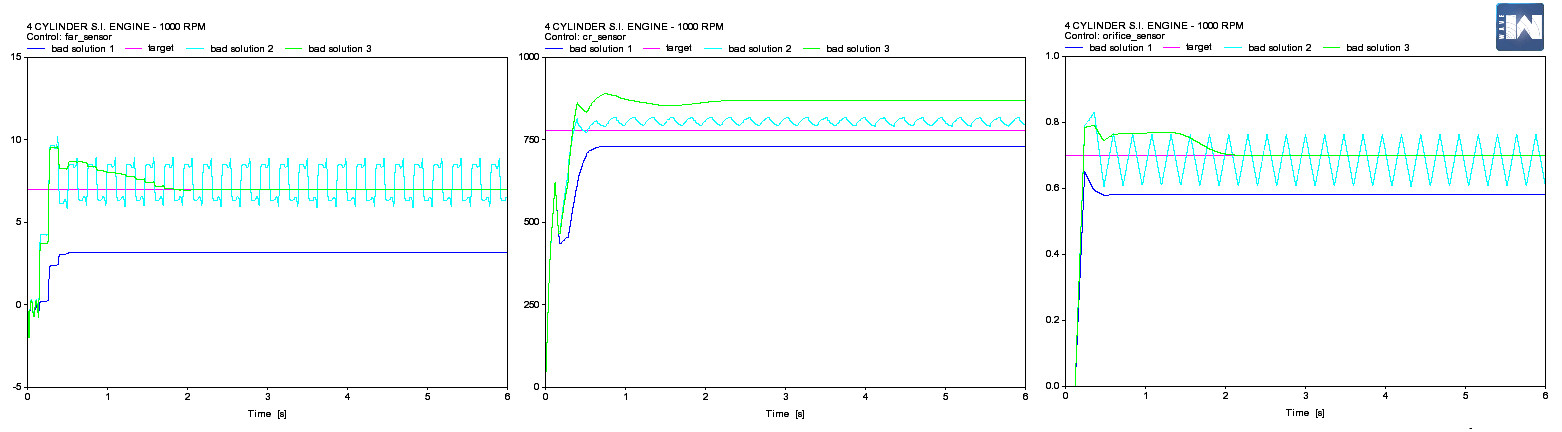}
\caption[The basic testing model: unsatisfactory solutions.]{The basic testing WAVE model: unsatisfactory solutions still provide enough information to estimate a fitting shift.}
\label{fig:bad}
\end{figure}

However, most candidate solutions found in the tuning process are not acceptable: oscillatory, convergent to a different value in one or more criteria, or even divergent (see figure \ref{fig:bad}). In either case, it is important to observe the development of the quantities' values in order to determine the shift $t_0$ used in the objective function: we choose a time point just before the initially erratic curve starts to follow a trend. For this model of overall simulation time 6 seconds, the best time shift found is $t_0 = 1.2$ second. It seems fitting to set the total simulation time such that the shift makes up the first $20\%$. That is, when all the controlled quantities react with approximately the same speed as in this case. In general, we need to compromise between the slowest and quickest responses: cut out the uninteresting information but leave enough to guide the search. Larger $t_0$ means greater risk of leaving out important information but possibly faster search.

Setting $t_0 >0$ is specific to our problem and certainly cannot be used in all applications of PID controllers. Whenever large overshooting is an issue, e.g. due to circuit breakers, this approach is not applicable. In our case, however, we have an empirical guarantee that leaving out most of the transient part (i.e. when $0< \tau < t_0$) does not lead to unacceptable solutions.
For the purposes of optimization, it is even desirable to ignore it as that reduces the ``noise'' in the objective function caused by this information. The problem then becomes easier to solve for the optimization method.

\subsection{Tuning CMA-ES} \label{ss:results}

The basic testing model was used to tune CMA-ES and other attributes. For that purpose, we consider all three controllers to be only PI controllers (i.e. with derivative gains set to zero) as is common in our application. 
Because the method uses randomness and each run is different, each test is run ten times, with no change to the given reference point. Then the minimum, maximum and average run times are computed.
The same reference point (i.e. scaling) of rather `poor quality'' is used for all tests. When we compare it to the vector of a good solution, the elements are off by two orders of magnitude on average.

A run is terminated when the target objective function value is hit or the budget of 12000 function evaluations is depleted. In each run, a different solution might be found, but setting the target value to $0.5$ ensures high quality of any of them (all the corresponding step responses are almost identical). 
In practice, however, the algorithm will have to be stopped manually by the user because the desired objective function value varies greatly from model to model and the target value is difficult to estimate.

The results are summarized in table \ref{t:all_parameters}. The best setting found is always compared to the setting, where one or two specified attributes were changed.



\vspace{10pt}

\begin{table}[H]
\scriptsize
\caption[]{ Experiments' results comparing various settings of CMA-ES and the objective function, 
						where always only the specified attribute or two were changed from the 
						BEST setting. 						
            BEST setting:     shift $t_0 = 1.2s$ (20\%), 
															CMA-ES variants = active and elitist, 
															adaptation method = CSA,
															population size $\lambda_0 = \lambda_{\text{def}}$, 
															parent set size $\mu_0 = 1/2 \lambda_{\text{def}}$, 
	                            restart parameter \texttt{TolFunHist} $= 1.5$.  
						Entries marked *: number of evaluations once exceeded the maximum number of iterations allowed. 
					  When computing the corresponding average value, the value of 12000 was used in such cases, 
					  resulting in lower-bound for the average run time.
					}
\label{t:all_parameters}
\centering
\renewcommand{\arraystretch}{1.4}
\begin{tabular}{|l|l||c|c|c|}
 \hline
 \multicolumn{2}{|l||}{setting}  & min & max  & average \\
 \hline
 \hline

 CMA-ES variants      & active, not elitist       & 644  & 10877      & 4145 \\
                      \cline{2-5}
                      & elitist, not active       & 327  & 5361       & 2184 \\
                       \cline{2-5}
                      & not active, not elitist   & 812  & $>$ 12000* & $>$ 5334 \\
 
 \hline
 
 adaptation method    & TPA                       & 926  & $>$ 12000* & $>$ 4724 \\

 \hline

 $\lambda_0, \mu_0$	  & $\lambda_0 = 2 \lambda_{\text{def}}, \ \mu_0 = \lambda_{\text{def}}$     
                                            & 1533 & 8812  & 3273 \\
		                  \cline{2-5}
                      & $\lambda_0 = 2 \lambda_{\text{def}}, \ \mu_0 = 1/2 \lambda_{\text{def}}$ 
										                        & 697  & 10973 & 4181 \\

 \hline
 \texttt{TolFunHist}  & $0.8 \times 1.5$    & 987  & 8980  & 3147 \\
                      \cline{2-5}
                      & $1.2 \times 1.5$    & 674  & 4990  & 2058 \\
                      \cline{2-5}
                      & $2.0 \times 1.5$    & 871  & 4960  & 2837 \\

 \hline
 shift $t_0$          & 0 s                       & 919  & 9567       & 3687 \\
											\cline{2-5}
											& 0.6 s (10\%)              & 354  & 11169      & 2963 \\
 \hline
 \hline
 
 \multicolumn{2}{|l||}{ BEST setting }      & 268 & 2267 & 1098    \\
	
 \hline

\end{tabular}
\end{table}


When choosing a variant of CMA-ES, the results show that the best combination is the elitist parent selection scheme with active updates of the covariance matrix and CSA step size adaptation (the Cumulative Step-size Adaptation [\citenum{hansenDerandomized}] uses information of the algorithm's overall progress across generations). The alternative TPA (Two-Point step-size Adaptation, [\citenum{cmaestpa}]) implements a line search along the direction of the latest mean shift.

Then we experiment with the initial population size $\lambda_0$ (i.e. the population size before the first restart, which is further used to compute the population sizes in the next ones). 
Using smaller than default populations is discouraged by Hansen [\citenum{CMAEStutorial}], while more exhaustive search with bigger populations motivated the development of the IPOP restart strategy. Therefore, we try to double the initial population size: $\lambda_0 = 2 \lambda_{\text{def}}$. 
The parent set size $\mu_0$, by default equal to half of the population size, is then either doubled as well (i.e. $\mu_0 = \frac{1}{2} \lambda_0$) or it stays the same (i.e. $\mu_0 = \frac{1}{2} \lambda_{\text{def}}$). 
For a given cost, smaller populations enable more generations than large populations, causing faster adaptation [\citenum{hansenDerandomized}], which seems beneficial in our application.

The best working value of the influential restart parameter \texttt{TolFunHist} (tolerance in function value history) for the basic testing model was found to be $n/2$, where $n=3$ is the number of controlled quantities. It was further observed that setting $n/2$ seems to be almost universally usable in our context, even though its optimal value differs for each model. 
The tolerance in function value was set to $\texttt{TolFun} = 0.1$ for the basic testing model and later roughly adjusted for other models, scaled in proportion to the ``usual'' numerical values of the objective function and expected threshold. 

In the objective function, shift $t_0 = 1.2$ second proved to be most effective.


\subsection{Testing robustness of the method}

Now we compare the tuned method's performance with various reference points. We take the baseline solution and derive the other reference points by multiplying \emph{each} of its elements by factors $10^{-3}, 10^{-2}, 10^{-1}, 10^1$, and $10^2$ (scaling by $10^3$ proved to be too challenging for all tested methods). These settings simulate user estimates of various quality.

In these tests, both PI and PID controllers are considered. Since we do not have a PID baseline solution with nonzero D parameters, we estimate them based on the corresponding magnitudes of P and I parameters.

Results are summarized in table \ref{t:all} together with other methods' robustness tests. We can see that the run times are very good: of 120 runs, 8 reached 2000 to 3000 evaluations and only two exceeded 3000 evaluations.  Surprisingly, there is no greater difference between the 6-dimensional PI setting and 9-dimensional PID setting. Higher problem dimension is balanced by greater flexibility of the system. 
Closer look at the behavior of the algorithms suggests that too large search area (i.e. scaling) causes them to ``get lost'' very far from the optimum. The run times are also influenced by the fact that there exist good solutions at the ``$10^{-1}$ level''.

We can conclude that when the reference point (scaling) is within a reasonable range of two orders of magnitude in each coordinate, the average run times are very usable. A user should be able to provide such a reference point -- an estimate of the solution. If in doubt, magnitudes of the reference point's coordinates should be chosen rather smaller than larger.

\begin{sidewaystable}[ph!]
\scriptsize
\caption{Results of algorithm testing on the basic model. 5 PSO runs, 5 SHADE runs and 10 CMA-ES runs were performed for each of 13 reference points. The value of ``-'' means that a satisfactory solution (i.e. solution with function value less than 0.5) was not found within the provided budget of 10000 function evaluations. Average run time was not computed if one or more runs did not finish within the given budget.}
\label{t:all}
\centering
\renewcommand{\arraystretch}{1.4}
\begin{tabular}{|l||c|c|c|c|c||c|c|c|c|c||c|c||c|c|c|}
 \hline
                    & \multicolumn{5}{|c||} {PSO runs}  & \multicolumn{5}{|c||} {SHADE runs}  & \multicolumn{2}{|c||}{CMA-ES runs} & \multicolumn{3}{|c|} {average run time}\\
 \cline{2-16}
 reference p.       & \#1 & \#2 & \#3 & \#4 & \#5       & \#1  & \#2  & \#3  & \#4  & \#5     & \phantom{.} min \phantom{.}  & \phantom{..}max\phantom{..}     & PSO  & SHADE & CMA-ES \\
 \hline
 \hline
 PI baseline        & 113 & 61 & - & - & -              & 130  & 396  & 341  & 2100 & 155     & 12  & 168     & -    & 624   & 76   \\
 \hline
 $10^{1}$ PI b.     & -   & -  & - & - & -              & - & - & - & - & -                   & 540 & 2064    & -    & -     & 1049 \\
 \hline
 $10^{2}$ PI b.     & -   & -  & - & - & -              & - & - & - & - & -                   & 821 & 5700    & -    & -     & 2061 \\
 \hline
 $10^{-1}$ PI b.    & 56  & 76 & 48  & 39  & 32         & 342  & 156  & 1560 & 260  & 538     & 61  & 816     & 50   & 571   & 317  \\
 \hline
 $10^{-2}$ PI b.    & 893 & -  & 483 & 226 & 403        & 5155 & 1514 & 2783 & 1437 & 2123    & 222 & 1334    & -    & 2602  & 592  \\
 \hline
 $10^{-3}$ PI b.    & -   & -  & -   & -   & 1600       & 4118 & 1090 & 2102 & 1243 & 1422    & 259 & 1617    & -    & 1995  & 812  \\
 \hline
 \hline
 PID baseline       & 562 & 28 & 105 & 71 & 28          & 428  & 307  & 497  & 441  & 658     & 32  & 256     & 159  & 466   & 67   \\
 \hline
 $10^{1}$ PID b.    & -   & -  & -   & -  & -           & - & - & - & - & -                   & 62  & 1462    & -    & -     & 1102 \\
 \hline
 $10^{2}$ PID b.    & -   & -  & -   & -  & -           & - & - & - & - & -                   & 953 & 4130    & -    & -     & 2022 \\
 \hline
 $10^{-1}$ PID b.   & 33  & 164 & 104 & 280  & 43       & 239  & 418  & 209  & 437  & 390     & 141 & 782     & 125  & 339   & 343  \\
 \hline
 $10^{-2}$ PID b.   & 387 & 624 & 301 & 1186 & -        & 6524 & 1018 & 1150 & 3297 & 1556    & 202 & 941     & -    & 2709  & 580  \\
 \hline
 $10^{-3}$ PID b.   & -   & -  & 1362 & 1734 & 1680     & 1673 & 2125 & 1945 & 1477 & 2163    & 416 & 2324    & -    & 1877  & 1138 \\
 \hline
 \hline
  calibration ref. p. & -   & -  & -    & -    & -        & - & - & - & - & -                   & 268 & 2267    & -    & -     & 1098 \\
 \hline
\end{tabular}
\end{sidewaystable}


\section{Comparison with PSO and SHADE} \label{s:comparison}

Performance of the above-described variant of CMA-ES on the problem of tuning coupled controllers was further compared with performance of two other prominent evolutionary algorithms: the particle swarm optimization (PSO) and the differential evolution (DE) in its success-history based adaptive variant SHADE.

\subsection{PSO}

PSO [\citenum{swarm1,swarm3}] is inspired by bird flocks or fish schools, where every individual moves by itself, yet the whole self-organized system acts as a single organism. Each particle within a swarm moves in the search space as it is assigned a different ``velocity'' vector in each generation. This vector is defined as a combination of the previous ``velocity'', the individual's best known position so far and the swarm's (or sub-swarm's) best known position. This way, the whole swarm moves towards historically best areas. 

For PSO implementation, we use the default setting of PSO in the DEAP (Distributed Evolutionary Algorithms in Python) optimization framework [\citenum{deap}]. We only add the essential scaling of variables same as in CMA-ES.

\subsection{SHADE}

The basic DE algorithm [\citenum{de2,de3,de1}] maintains a population of candidate solution, which are further combined and tested for better objective function values. Each candidate solution $X$ is tested against a new point $Z$, which was obtained as a binary crossover of $X$ and $Y = A + f (B-C)$, where $A$, $B$ and $C$ are three distinct candidate solutions in the population and $f \in [0,2]$ is a parameter. The crossover probability of each vector element is given by parameter $cr \in [0,1]$. 
Both parameters and the population size greatly influence efficiency of the search.

Considering the context of our problem and the performance-comparing tests in [\citenum{shade_budget}], the SHADE variant of DE was chosen to compete with CMA-ES. The SHADE upgrade [\citenum{shade}] introduces adaptation of the DE parameters $cr$ and $f$ based on success history that is stored in an external archive.

Implementation of the SHADE algorithm by Tanabe was used [\citenum{shade_implementation}] and tuned in accordance with its author's recommendations [\citenum{shade_budget}] and additional experiments, setting, most notably, initial $cr = 0.5$, initial $f = 0.5$ and population size $= 2N$, where $N$ is the problem dimension. One restart criterion was added to the standard SHADE algorithm, based on recommendations in [\citenum{shade_budget}]: if the number of objective function evaluations exceeds 1000, restart can be launched if the best-so-far solution is not updated for $50 \times N$ evaluations. The particular numbers were set based on experiments and the first condition proved very important.

Unlike in CMA-ES and PSO, the reference point is not used for scaling but to define boundaries of the area, from which the initial population is sampled. Each coordinate of the reference vector is multiplied by $-10$ and $10$ to define the boundaries. Candidate solutions in further generations are not restricted.

\subsection{PSO and SHADE experiment results}

The experiments with PSO and SHADE were performed with same set of 13 reference points as in the previous tests of CMA-ES, with 5 runs of each test. The stopping criteria were either hitting the target value of the objective function or depleting the budget of $10000$ objective function evaluations. All test results can be found in table \ref{t:all}.

PSO was able to outperform CMA-ES but was not very reliable. Its good performance was limited to very good reference points and the algorithm very often (36 out of total 65 runs) did not converge within the given budget.

The performance of SHADE was more consistent but very slow compared to CMA-ES. Same as PSO, SHADE never converged within the given budget for the reference points of baselines multiplied by $10^1$ and $10^2$ or the one used for CMA-ES calibration.


\section{Verification of the method on models of real engines} \label{s:rw_models}

We have shown that, on the testing model, the described version of CMA-ES performs very well and clearly outperforms PSO and DE. Now we verify its usability on models of real engines provided by Ricardo. The testing set consists of two models with 1 controller (labeled M1.1, M1.2), three models with 2 controllers (labeled M2.1, M2.2, M2.3) and one model with 3 controllers (labeled M3.1). 
For each, we are given a baseline setting of controllers: a setting tuned manually by an engineer, which is always in the PI form (i.e. with D gain set to zero). When a baseline value of the D gain is needed in our tests, we estimate its magnitude from the P and I gains.

The single-controller models are tested with 5 reference points: the baseline solution and its element-wise multiples by factors $10^{-2}, 10^{-1}, 10^1$ and $10^2$. The multi-controller models are tested with three reference points: the baseline solution and its multiples of $10^{-1}$ and $10^1$. The primary reason for this restriction is the enormous time consumption. Each test is run five times. 
A run is terminated upon reaching the (empirically set) target value specific to the given model. Even though each run might produce a different solution, as long as the target value is reached, their quality is guaranteed.

All results are summarized in table \ref{t:rwm}.


\vspace{10pt}

\begin{table}
\scriptsize

\caption[]{Real-world models}
\label{t:rwm}
\centering
\renewcommand{\arraystretch}{1.4}

\begin{tabular}{p{6.5cm} p{6.5cm}}

\begin{minipage}{.5\linewidth}

\begin{tabular}{|c|l||c|c|c|}
 \hline
 model & reference p.    & min  & max   & aver. \\
 \hline
 \hline 
       & PI baseline           & 2 & 68 & 28 \\
       \cline{2-5}
       & $10^{1}$ PI b.        & 35 & 153 & 79 \\
       \cline{2-5}
 M1.1  & $10^{2}$ PI b.        & 95 & 519 & 225 \\
       \cline{2-5}
       & $10^{-1}$ PI b.       & 20 & 120 & 66 \\
       \cline{2-5}
       & $10^{-2}$ PI b.       & 49 & 296 & 123 \\
 \hline
 \hline  
       & PI baseline           & 1 & 22 & 9 \\
       \cline{2-5}
       & $10^{1}$ PI b.        & 4 & 28 & 11 \\
       \cline{2-5}
  M1.2 & $10^{2}$ PI b.        & 80 & 225 & 187 \\
       \cline{2-5}
       & $10^{-1}$ PI b.       & 34 & 100 & 51 \\
       \cline{2-5}
       & $10^{-2}$ PI b.       & 57 & 181 & 94 \\
 \hline
\end{tabular}

\vspace{74pt}

\end{minipage} 

& %

\begin{minipage}{.5\linewidth}

\begin{tabular}{|c|l||c|c|c|}
 \hline
 model & reference p. & min  & max   & aver. \\
 \hline
 \hline 
       & PI baseline           & 11  & 66  & 35 \\
       \cline{2-5}
 M2.1  & $10^{1}$ PI b.        & 244 & 280 & 255 \\
       \cline{2-5}
       & $10^{-1}$ PI b.       & 4   & 32  & 21 \\
 \hline
 \hline 
       & PI baseline      & 8 & 98 & 29 \\
       \cline{2-5}
 M2.2  & $10^{1}$ PI b.   & 60 & 770 & 364 \\
       \cline{2-5}
       & $10^{-1}$ PI b.  & 44 & 107 & 64 \\
 \hline
 \hline 
       & PI baseline      & 9 & 78 & 32 \\
       \cline{2-5}
 M2.3  & $10^{1}$ PI b.   & 250 & 757 & 629 \\
       \cline{2-5}
       & $10^{-1}$ PI b.  & 49 & 1188 & 347 \\
\hline
       \hline 
       & PID baseline     & 10 & 91 & 57 \\
       \cline{2-5}
 M2.3  & $10^{1}$ PID b.  & 274 & 857 & 522 \\
       \cline{2-5}
       & $10^{-1}$ PID b. & 82 & 1576 & 749 \\
 \hline
 \hline 
       & PID baseline      & 41  & 331  & 152 \\
       \cline{2-5}
 M3.1  & $10^{1}$ PID b.   & 827 & 1763 & 1268 \\
       \cline{2-5}
       & $10^{-1}$ PID b.  & 179 & 3867 & 2476 \\ 
 \hline
\end{tabular}

\end{minipage} %

\end{tabular}

\end{table}


\subsection{Single-controller models}

Even though our primary objective is to tune multiple controllers, we include also single-controller models as they are very important in practical use and we want our method to work for them as well. 
Model M1.1  contains a turbocharger and M1.2 contains a twin turbocharger. They represent the typical use of a PI controller in a combustion engine. 

With the model dimension being only 2, the run times for both M1.1 and M1.2 are very short. Good reference points for single controllers can be obtained by commonly used rules of thumb (e.g. Ziegler-Nichols) or provided by previous (personal or programmed) experience with similar models.


\subsection{Multi-controller models}

The two-controller models M2.1 and M2.2 do not pose a greater challenge than the single-controller models. 
For model M2.3, we compare PI and PID control. The extreme differences in minimal and maximal run time values of ``$10^{-1}$ PI baseline'' and ``$10^{-1}$ PID baseline'' tests are caused by non-optimal setting of the restart parameter tolerance in function value history ($\texttt{tolhistfun}$) described above. After adjustment, the ``$10^{-1}$ PI baseline'' and ``$10^{-1}$ PID baseline'' run times drop and the results of PI control resemble those of M2.2.

Unlike the previous cases, the three-controller model M3.1 requires a full PID control. The reference point was obtained by estimating the D gains and adding them to the given (not very good) PI engineer-tuned baseline ``solution''. It can be seen that this model is considerably harder to tune than the previous models. While the algorithm has no trouble finding the same near-optimal ``solutions'' similar to the one given by an engineer, it was hard to get to an actual global optimum, when all three controllers converge.


\section{Conclusion}

This paper has shown how the Covariance Matrix Adaptation Evolution Strategy can be applied to the problem of tuning the gains of multiple coupled PID controllers within combustion engine simulations. We have shown that its version with bi-population restart scheme, elitist parent selection and active covariance matrix updates is capable of finding good parameters of up to three PID controllers through minimization of a fitting objective function. The method has been calibrated on a testing model and verified on models of real-world engines, showing its practical usability and tolerable computation times even for poor-quality reference points. On the testing model, CMA-ES clearly outperformed PSO and SHADE methods.


\section*{Acknowledgments}

This research did not receive any specific grant from funding agencies in the public, commercial, or not-for-profit sectors. The author would like to thank Ricardo Prague, s.r.o. for their support and to anonymous reviewers, whose comments and suggestions inspired significant improvements.



\end{document}